\documentclass[
    ,final            % use final for the camera ready runs
  ]
  {aipproc}

\layoutstyle{6x9}

\begin{document}

\title{Recent Developments in the Lorentz Integral Transform (LIT) Method}

\author{G.Orlandini}{
  address={Dipartimento di Fisica, Universit\`a di Trento and INFN, Gruppo
  Collegato di Trento,\\ Via Sommarive 14, I-38050 Povo, Italy}
}
\begin{abstract} Recent results on electromagnetic (e.m.) reactions
into the continuum of systems with A from 3 to 7 are presented.
They have been obtained using the LIT method \citep{ELO94}.
The method is shortly reviewed, emphasizing how all the results, 
though obtained with the sole ingredient of the N-N potential, contain 
the full complicated dynamics of the A-body system, both in the initial 
and in the final states. 
\end{abstract}

\maketitle

%%%%%%%%%%%%%%%%%%%%%%%%%%%%%%%%%%%%%%%%%%%%
%% MAINMATTER
%%%%%%%%%%%%%%%%%%%%%%%%%%%%%%%%%%%%%%%%%%%%
\section{The LIT method}

The LIT method is suited for {\it ab initio} studies of few 
(many) body dynamics in the framework of non relativistic
 quantum mechanics.
The inputs are the nuclear potential and the excitation operator 
relevant to the reaction. Even if the reaction brings the system to
a continuum state one does not need to calculate continuum wave functions.
What the method allows to calculate are matrix elements (or combinations 
of them) to continuum states, which is what one needs in order 
to evaluate cross sections.

Essential to the method are solutions of the Schr\"odinger equation
for the ground state and of a Schr\"odinger-like equation with a
source. In both cases the solutions have bound-state asyntotic 
conditions. So in order to apply 
the LIT method one only needs a "good" technique for bound 
state calculations.

The LIT method
applies both to inclusive reactions and to exclusive ones. In 
a traditional approach the former case is considered more 
complicated, because of the necessity to know solutions
for all the disintegration channels. However, it turns out that
the application of the LIT method is more straightforward in
this case than in an exclusive reaction.
The LIT method has been benchmarked in 2- and 3-body systems where 
continuum states can be calculated directly \cite{ELO94,
Lapiana,Martinelli,Golak}

The procedure of the LIT method, though similar in spirit, is different
for inclusive and  exclusive reactions. In the following I will list
the main steps one has to take in the two cases.

\subsection{Inclusive Reactions}
For inclusive e.m. reactions one needs to know the so called
response function
\begin{equation}
R(\omega)=\sum\!\!\!\!~\!\!\!\!\!\!\!\!\int _n|\langle 0|\Theta|n\rangle|^2\delta(\omega-E_n+E_0)
\end{equation}
where $\omega$ represents the energy transferred by the elctromagnetic 
probe,  $|0>$ and
$E_0$ are ground state wave function and energy of the system undergoing 
the reaction, $|n>$ and $E_n$ are all the eigenstates and eigenvalues of 
the Hamiltonian $H$ and $\Theta$ is the operator relevant to the
reaction. In order to calculate $R(\omega)$ one proceeds in three steps.

{\it Step 1.} 
The first step consists in solving for many $\omega_0$ and 
fixed $\Gamma$ the following equation 
\begin{equation} 
(H- E_0-\omega_0 + i \Gamma)\tilde\Psi = \Theta |0>
\end{equation}
The values of the parameters $\omega_0$ and $\Gamma$ are chosen in
relation to the physical problem. For example, if one wants
to calculate the cross section in a range of energies from 0 to 100 MeV
one solves that equation with $\Gamma\simeq $10 MeV for about 100 
$\omega_0$ values chosen 
in a range slightly larger than 100 MeV, e.g. $100+2\Gamma$ MeV.
As it will be clear below 
the value of $\Gamma$ is related to the kind of resolution one 
wishes to have in order to to resolve the expected structures 
of the response function and the values of $\omega_0$ scan the region
of interest.

{\it Step 2.} The second step consists in calculating the overlaps 
$<\tilde\Psi|\tilde\Psi>$ of the solutions. Of course these overlaps
depend on $\omega_0$ and $\Gamma$. A theorem based on the closure property
of the Hamiltonian eigenstates ensures that this dependence can be 
expressed as
\cite{ELO94}
\begin{equation}
 <\tilde\Psi|\tilde\Psi>= \int R(\omega)\, L(\omega,\omega_0,\Gamma)\, 
d\omega
\end{equation}
where $L(\omega,\omega_0,\Gamma)$ is the Lorentzian function centered at
$\omega_0$ and with $\Gamma$ as width.
Therefore if one solves Eq. (1) one can easily obtain the Lorentz integral
transform of the response function. 

{\it Step 3.} The third step consists in the inversion of this transform 
in order to obtain the response function.

\subsection{Exclusive Reactions}

For exclusive reactions the LIT method is applied 
\cite{Lapiana,Quaglioni} to calculate the 
relevant transition matrix element $<\Psi (E)|\Theta|0>$, 
where $\Psi (E)$ is the wave function in the continuum at energy E.
As already said, in this case, the procedure requires more steps.

{\it Step 1.} The first step is identical to the case of inclusive 
reactions i.e. one has to solve Eq.(1). Now the choice
of the parameters is dictated by different criteria.
I will comment on this after describing  the last step. 

{\it  Step 2.} There is a second Schr\"odinger like equation with source 
to be  solved for exclusive reactions. For example, in the case of a 
2-body break-up reaction this is 
\begin{equation}
(H- E_0-\omega_0 + i \Gamma)\tilde\Psi_2 = V |0>
\end{equation}
where V is the potential acting between the particles belonging 
to the two different fragments.

{\it Step 3.} The third step consists in calculating the overlap 
$<\tilde\Psi|\tilde\Psi_2>$.
Again, for a theorem based on the closure property
of the Hamiltonian eigenstates, this overlap is connected via a LIT to 
the important function $F(E)$ \cite{Lapiana,efros85}
\begin{equation}
 <\tilde\Psi|\tilde\Psi_2>= \int F(E)\, L(E,\omega_0,\Gamma)\, 
dE
\end{equation}
As it will be clear in step 5 this function is what one needs to calculate
the matrix element of interest.

{\it Step 4.} In order to obtain F(E) one has to invert the previous 
Lorentz transform and proceed to the final step.

{\it Step 5.} It is easy to show \cite{Lapiana, efros85} that the matrix 
element of interest is connected to the function $F(E)$ via the following
integral 

\begin{equation}
<\Psi(E)| \Theta|0> = \int dE'\, {F(E')\over (E-E'+i\epsilon)}
\end{equation}
This equation suggests the physical criteria one has to follow in
the choice of
the parameters $\omega_0$ and $\Gamma$ entering Eqs.(1) and (4).
Since the reaction matrix element is taking its major contribution from
$F(E')$ at $E'=E$ a good knowledge of $F(E')$ is needed around
the $E$ value. This means that its Lorentz transform needs to be
known in a large enough range around that value and for small enough 
width in order to obtain an accurate function when inverting it.

Important remarks regarding the solutions of  Eqs.(1) and (4) and about the
inversion of the LIT will be done in the next section.

\subsection{Important Remarks}

At this point it should be clear that the main points of the LIT method 
lay in Eqs. (1) and (4). One can easily show that the solution of these 
equations is unique. Moreover a very important point is that, due to the 
presence of a complex energy and sources which asyntotically vanish the 
solutions are limited i.e. of bound state type (this is 
certainly the case for all physical e.m. operators and for the nuclear 
potential V in Eq.(4); for long range potentials a different procedure is 
requred \cite{Lapiana}). This has the very 
important consequence that one only needs bound state techniques (notice 
that also the ground state is an input of Eqs. (1) and (4)) in order to
solve the problem of calculating a reaction in the continuum, even when 
many complicated channels are opened, like for inclusive cases above
the thresholds. 

An other remark concerns the inversion of the transforms. Due to the
bell shaped form of the kernel, the inversions turn out to be rather 
stable. Of course for any inversion technique one can use 
(we use the regularization method \cite{book}) one has to check that
the same result for different choices of $\omega_0$ sets and
values of $\Gamma$ is obtained.

In the following I will present results which have been obtained 
using the Correlated Hyperspherical Harmonics (CHH \cite{CHH})
and the Effective Interaction in the Hyperspherical Harmonics 
expansions (EIHH \cite{EIHH}) techniques to solve
the bound sate problems and both semirealistic (MTI-III \cite{MT} and AV4'
\cite{AV4'}) and realistic (AV18+UIX \cite{AV18UIX}) potentials.

\section{results}

I start showing results for the 3-body systems \cite{3body}. In Fig.
1a the longitudinal responses of $^3$H and $^3$He are shown. 
The effect of 
the 3-body force is visible in the quasi elastic (q.e.) peak. While the 
3-body potential  seems to be necessary for a better agreement with data 
in the case of $^3$He, it brings the curves farer from them in the tritium
case. The comparison with data at higher momentum transfers $q$ 
 is of the same quality \cite{3body}. 

\bigskip
%%%%%%%%%%%%%%%%%%%%%%%%%%%%%%%%%%%%%%%%%%%%%%%%%%%%%%%%%%%%%%%%%%%%%%%

\begin{figure}[htb]
\begin{minipage}[t]{0.47\textwidth}
%\center{(a)}
\includegraphics[height=.32\textheight,width=\textwidth]{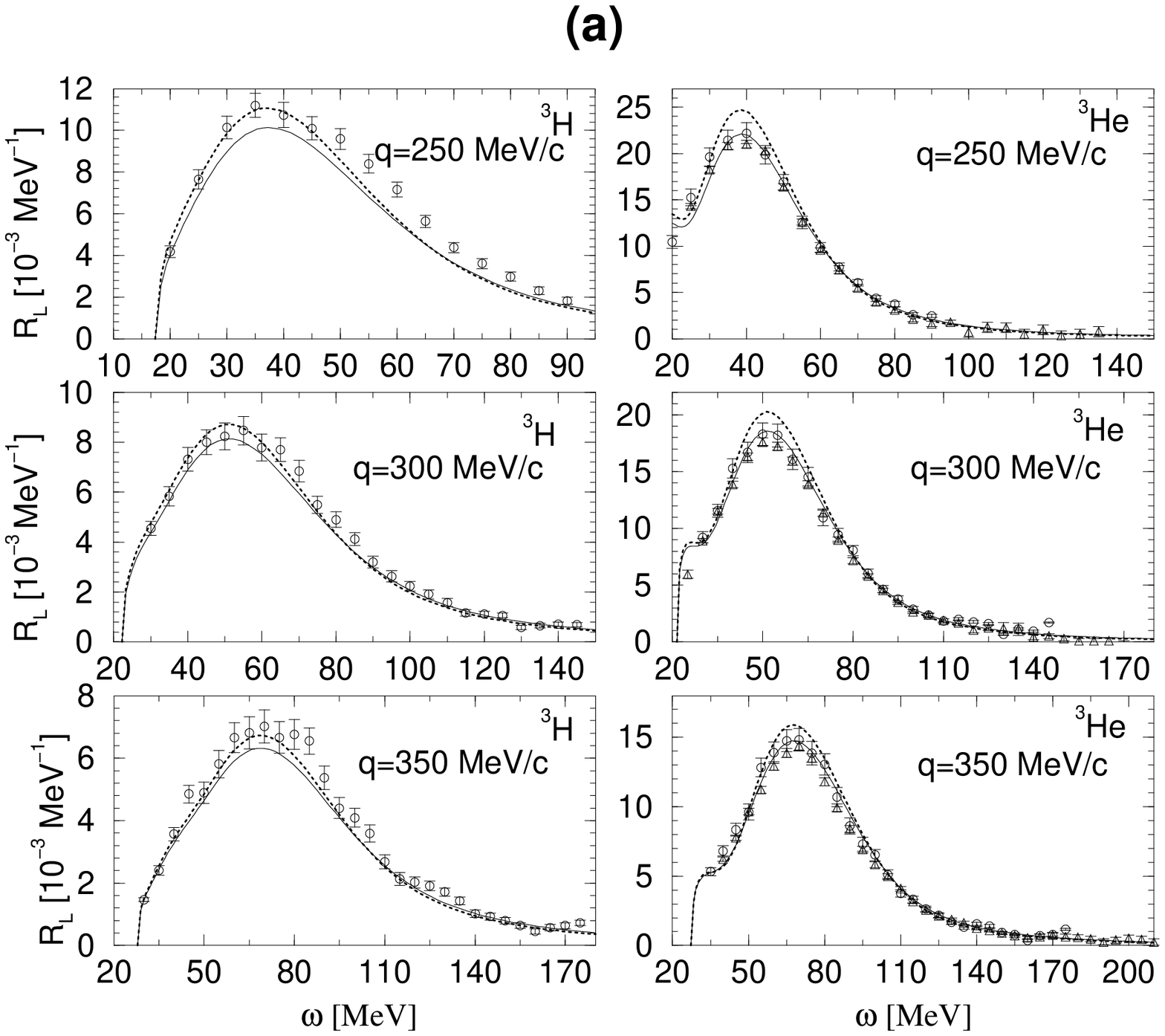}
%\caption{Faddeev (dots) and LIT result (upper and lower bounds, 
%full curves) in unretarded E1 approximation and for AV18 potential. 
%From   \cite{bench02}.}
\end{minipage}
\hspace{\fill}
\vspace{-1cm}
\begin{minipage}[t]{0.47\textwidth}
%\center{(b)}
%\begin{minipage}[t]{75mm}
%\framebox[74mm]{\rule[-26mm]{0mm}{52mm}}
\includegraphics[height=.30\textheight,width=\textwidth,angle=0]
{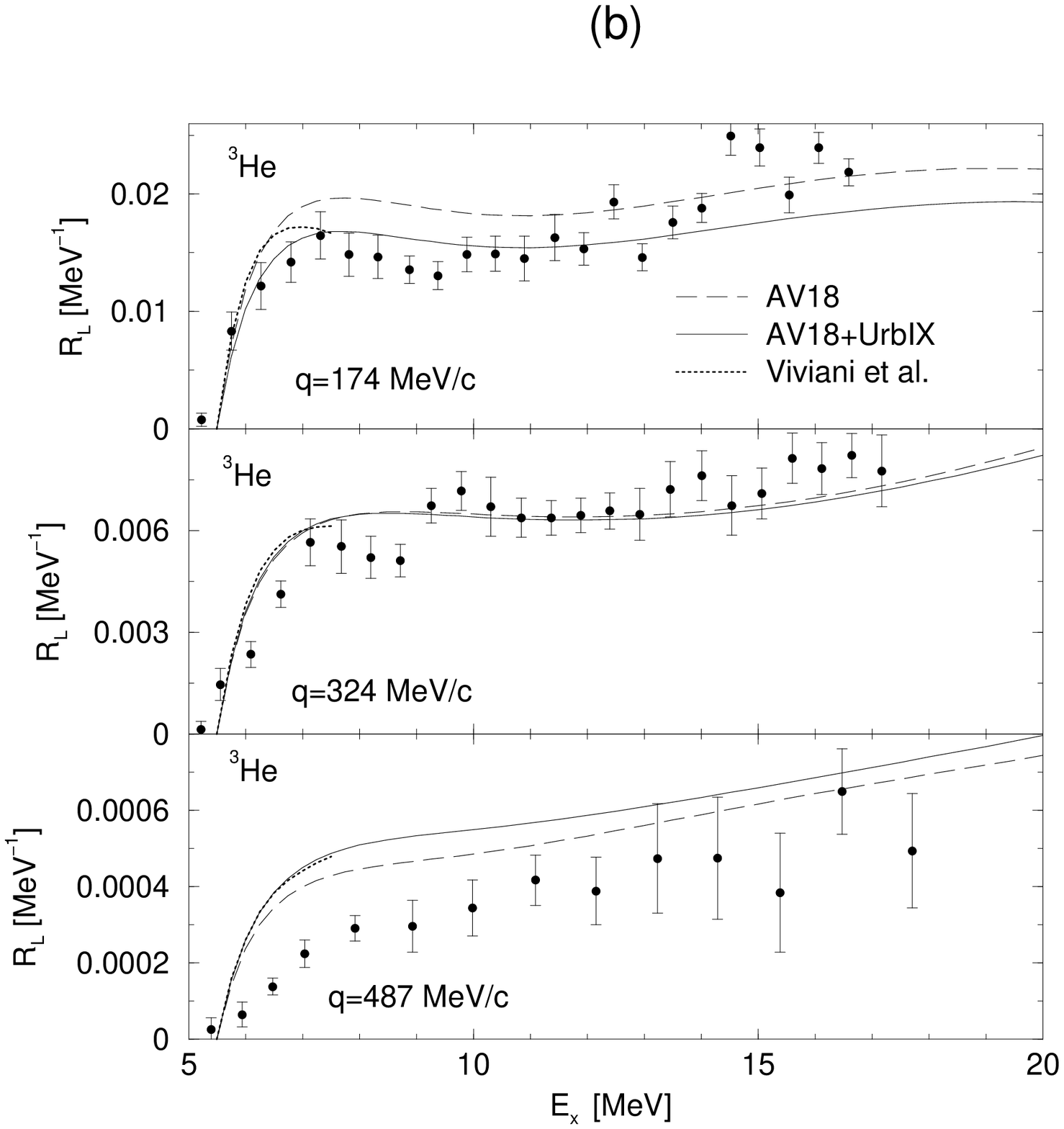}
\caption{Longitudinal electron scattering responses of 3-body systems. 
LIT results in unretarded E1 approximation with AV18 only (dashed curves) 
and with AV18+UIX (full curve).(a): q.e. region, data from 
\cite{saclay,bates}. (b): low energy region, data from \cite{retzlaff},
dotted curve: result from \cite{Viviani}}
\end{minipage}
\end{figure}
%%%%%%%%%%%%%%%%%%%%%%%%%%%%%%%%%%%%%%%%%%%%%%%%%%%%%%%%%%%%%%%%%%%%%%%

Fig. 1b
shows the situation at lower energies. The discrepancy at the highest $q$
is worsened by considering the 3-body interaction.
More precise data would be needed to study 3-body force effects.

An interesting effect is shown in Fig. 2 where the frame dependence of the
results is shown. Performing the calculation in the laboratory system, 
(where the total momentum of the initial state is at rest) or 
in the anti-laboratory one (where the total momentum of the
final state is at rest) or in the Breit
frame (where the total momenta of the initial and final states are 
equal to $q/2$) brings to a visible shift of the q.e. peak at higher 
momenta.
This shows the limits of the non realtivistic calculation. Data seem to
agree better with the results obtained in the Breit frame.

In Table 1 results for the longitudinal 2-body electrodisintegration 
of $^4$He are shown \cite{Quaglioni} for the kinematics of 
Ref.\cite{ducret} represented 
in the ($q-\omega$) plane and labelled by arabic numbers in Fig 3. 
The different effects of proper
antisymmetization (AS) and of the interaction in the final state (FSI), 
neglected in the direct knock-out plane wave approximation (PWIA), 
are shown. 
In general one notices small effects of the antisymmetrization 
for almost all cases, as one would expect in parallel kinematics.
Nevertheless for the kinematics at lower energies these effects
can increase up to about 10\%. The role of FSI is much more important,
especially at low $q$. It seems to decrease considerably at $q$
beyond 500 MeV/c. All these results have been obtained with the 
semirealistic MTI-III potential (more details can be found in
the poster contribution to this conference \cite{sofia}). 
Results for the 4-body system
with AV18+UIX potential have been obtained in the inclusive case
(total photodisintegration) and are also shown as poster contribution to 
this conference \cite{doron}

%%%%%%%%%%%%%%%%%%%%%%%%%%%%%%%%%%%%%%%%%%%%%%%%%%%%%%%%%%%%%%%%%%%%%%%
\vspace{2mm}
\begin{figure}[htb]
\includegraphics[width=.44\textwidth,height=.33\textheight]{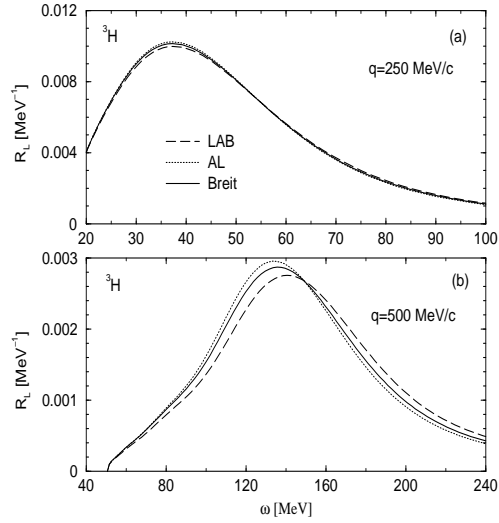}
\caption{Frame dependence of the longitudinal electron scattering 
responses of triton calculated in the laboratory system (LAB), or
in the anti-laboratory one (AL) or in the Breit frame.}
%\label{fig1}
\end{figure}
%%%%%%%%%%%%%%%%%%%%%%%%%%%%%%%%%%%%%%%%%%%%%%%%%%%%%%%%%%%%%%%%%%%%%%%

%%%%%%%%%%%%%%%%%%%%%%%%%%%%%%%%%%%%%%%%%%%%%%%%%%%%%%%%%%%%%%%%%%%%%%%
%\vspace{2mm}
%\begin{figure}[htb]
%\includegraphics[width=\textwidth,height=.22\textheight]{kinematics.eps}
%\caption{Frame dependence of the longitudinal electron scattering 
%responses of triton calculated in the laboratory system (LAB), or
%in the anti-laboratory one (AL) or in the Breit frame.}
%\label{fig1}
%\end{figure}
%%%%%%%%%%%%%%%%%%%%%%%%%%%%%%%%%%%%%%%%%%%%%%%%%%%%%%%%%%%%%%%%%%%%%%%
%%%%%%%%%%%%%%%%%%%%%%%%%%%%%%%%%%%%%%%%%%%%%%%%%%%%%%%%%%%%%%%%%%%%%%%
\vspace{2mm}
\begin{figure}[htb]
\includegraphics[width=.55\textwidth,height=.22\textheight]{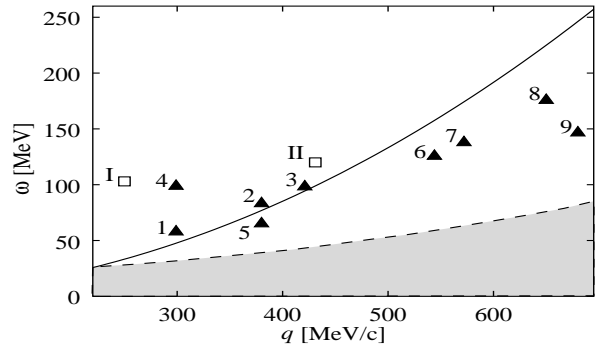}
\caption{Values of momentum $q$ and energy transfer $\omega$ for the
kinematics of Ref. \cite{ducret}. See Table 1.}
%\label{fig1}
\end{figure}
%%%%%%%%%%%%%%%%%%%%%%%%%%%%%%%%%%%%%%%%%%%%%%%%%%%%%%%%%%%%%%%%%%%%%%%
%%%%%%%%%%%%%%%%%%%%%%%%%%%%%%%%%%%%%%%%%%%%%%%%%%%%%%%%%%%%%%%%%%%%%%%

%\begin{figure}[htb]
%\begin{minipage}[t]{0.47\textwidth}
%\center{(a)}
%\includegraphics[height=.30\textheight,width=\textwidth]{Figure3a.eps}
%\caption{Faddeev (dots) and LIT result (upper and lower bounds, 
%full curves) in unretarded E1 approximation and for AV18 potential. 
%From   \cite{bench02}.}
%\end{minipage}
%
%\hspace{\fill}
%
%\vspace{-1cm}
%\begin{minipage}[t]{0.47\textwidth}
%\center{(b)}
%\begin{minipage}[t]{75mm}
%%\framebox[74mm]{\rule[-26mm]{0mm}{52mm}}
%\includegraphics[height=.28\textheight,width=\textwidth,angle=0]
%{Figure3b.eps}
%\caption{(a): Percentage effects of antisymmetrization ( Longitudinal $^4$He(e,e'p)$^3$H cross section}
%\end{minipage}
%\end{figure}
%%%%%%%%%%%%%%%%%%%%%%%%%%%%%%%%%%%%%%%%%%%%%%%%%%%%%%%%%%%%%%%%%%%%%%%
\begin{table}
\caption{Results for the longitudinal 2-body electrodisintegration 
response function $R_L$ of $^4$He. In the first 3 columns the kinematical
 values of momentum $q$, energy transfer $\omega$ and missing momentum 
$p_m$ from Ref. \cite{ducret}. The AS and FSI effects are shown as 
percentages of the PWIA values} 
\renewcommand{\arraystretch}{1.1}
\begin{tabular}{ccccccc}
 Kin. & $q$ & $\omega$ & $p_m$  & $R_L(PWIA)$ & 
 AS & AS+FSI \\
No. & [MeV/c] & [MeV] & [MeV/c] & [(GeV/c)$^{-3}$ sr$^{-1}$] & $\%$ & 
$\%$ \\
\hline
1 & 299 & 57.78 & +30 & 110.1 & +10.9 & -45 \\ 
2 & 380 & 83.13 & +30 & 83.7  & +2.0  & -25 \\
3 & 421 & 98.19 & +30 & 72.0  & +0.7  & -18 \\
4 & 299 & 98.70 & -90 & 61.6  & +1.9  & -35 \\
5 & 380 & 65.06 & +90 & 44.6  & +7.5  & -37 \\
6 & 544 & 126.6 & +90 & 23.2  & -0.1  & +0.6 \\
7 & 572 & 137.82 &+90 & 20.6  & -0.2  & +3.4 \\
8 & 650 & 175.67 &+90 & 14.7  & -0.2  & +7.3 \\
9 & 680 & 146.68 &+190 & 1.79 & -0.4  & +28.8 \\
\hline
\end{tabular}
\end{table}

%%%%%%%%%%%%%%%%%%%%%%%%%%%%%%%%%%%%%%%%%%%%%%%%%%%%%%%%%%%%%%%%%%%%%%%%
In Fig. 4a results for the total photodisintegration of the 
6-body systems are
shown \cite{6body}. The unretarded electric dipole operator is used.
A very interesting feature i.e. the appearance of two 
resonances in the case of $^6$He is worth to be noticed. They likely
correspond to the soft mode due to theoscillation of the two halo 
neutrons against the alpha-core and to the classical giant resonance
 mode of protons against neutrons,
respectively. Since the latter requires the break-up of the alpha core
the resonance  appears at higher energy. One finds a unique resonance 
in $^6$Li, though also in this case one would expect two, due
to the probable clusterized form of this nucleus. 
An explanation for 
the disappearence of the dip between the two resonances might be that 
in this case an additional mode i.e. the $^3$H-$^3$He oscillation
is filling the gap.
An analogous $^3$H-$^3$He mode in $^6$He would not be excited by 
the dipole operator. New accurate 2-body break up data 
of $^6$Li would be needed to confirm this hypothesis.
%%%%%%%%%%%%%%%%%%%%%%%%%%%%%%%%%%%%%%%%%%%%%%%%%%%%%%%%%%%%%%%%%%%%%%%

\begin{figure}[htb]
\begin{minipage}[t]{0.47\textwidth}
\center{(a)}
\includegraphics[height=.30\textheight,width=\textwidth]{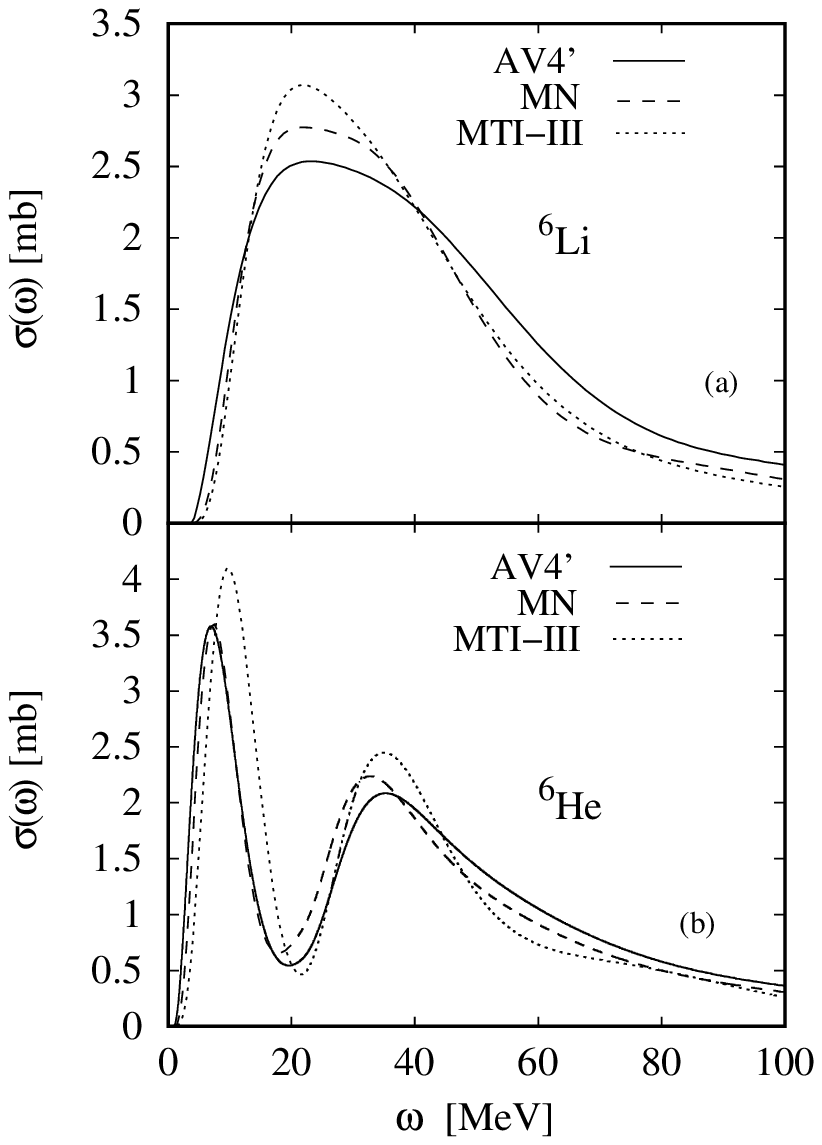}
\caption{Faddeev (dots) and LIT result (upper and lower bounds, 
full curves) in unretarded E1 approximation and for AV18 potential. 
From   \cite{bench02}.}
\end{minipage}
\hspace{\fill}
\vspace{-1cm}
\begin{minipage}[t]{0.47\textwidth}
\center{(b)}
%\begin{minipage}[t]{75mm}
%\framebox[74mm]{\rule[-26mm]{0mm}{52mm}}
\includegraphics[height=.31\textheight,width=\textwidth,angle=0]
{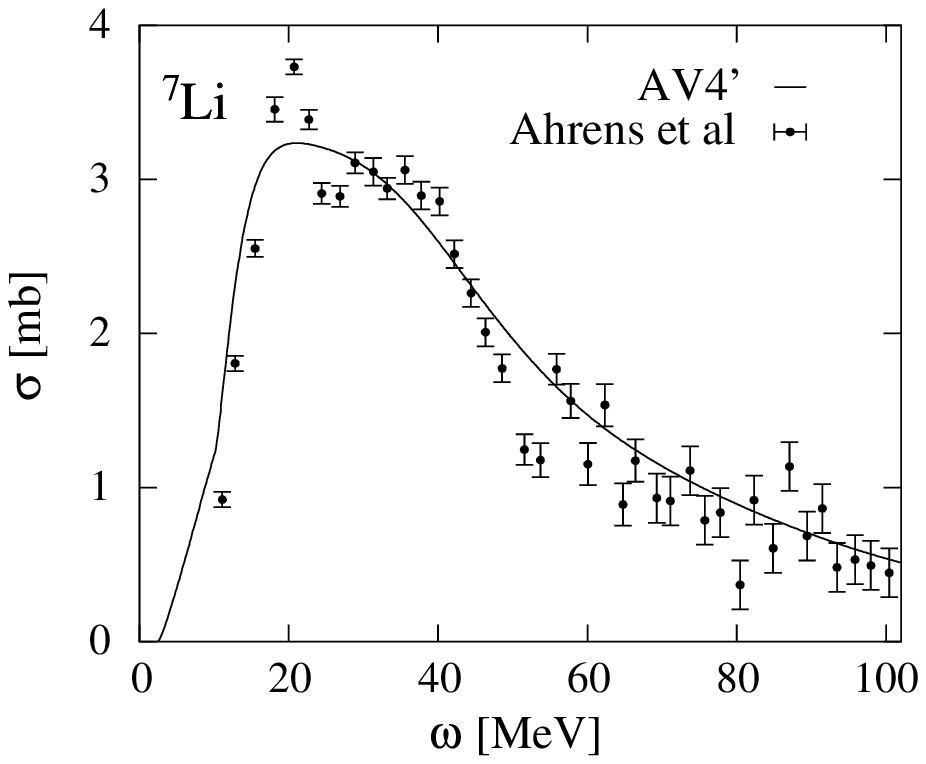}
\caption{(a): total photoabsorption cross sections of $^6$Li and $^6$He 
for different potential models \cite{MT,AV4',MN}. (b): total 
photoabsorption
cross section of $^7$Li compared to data of Ref. \cite{ahrens}.}
\end{minipage}
\end{figure}
%%%%%%%%%%%%%%%%%%%%%%%%%%%%%%%%%%%%%%%%%%%%%%%%%%%%%%%%%%%%%%%%%%%%%%%
Finally in Fig. 4b it is shown how the LIT method presently allows to 
calculate an inclusive reaction involving 7 particles fully 
{\it ab initio}. This result for the total photoabsorption of $^7$Li 
is commented in more detail in another contribution to this conference
\cite{7body}
%% If the bibliography is produced without BibTeX comment out the



\IfFileExists{\jobname.bbl}{}
 {\typeout{}
  \typeout{******************************************}
  \typeout{** Please run "bibtex \jobname" to optain}
  \typeout{** the bibliography and then re-run LaTeX}
  \typeout{** twice to fix the references!}
  \typeout{******************************************}
  \typeout{}
 }

\end{document}